\documentclass[]{iopart}
        \usepackage{ amsmathred, latexsym, amssymb, amscd,amsfonts, amsthm,epsfig,color,times,appendix}
\usepackage{graphicx}

\usepackage{pictex}
\usepackage[english]{babel}
\newtheorem{theorem}{\bf Theorem}[section]

\begin{document}

\title[Statistically efficient tomography of low rank states with incomplete measurements]{Statistically efficient tomography of low rank states with incomplete measurements}

\author{ \sf \bfseries  Anirudh Acharya$^{1}$, Theodore Kypraios$^{1}$, M\u{a}d\u{a}lin Gu\c{t}\u{a}$^{1}$} 
\address{$^{1}$School of Mathematical Sciences, University of Nottingham,
University Park, NG7 2RD Nottingham, United Kingdom}
\date{}

\begin{abstract}
The construction of physically relevant low dimensional state models, and the design of appropriate measurements are key issues in tackling quantum state tomography for large dimensional systems. We consider the statistical problem of estimating low rank states in the set-up of multiple ions tomography, and investigate how the estimation error behaves with a reduction in the number of measurement settings, compared with the standard ion tomography setup. We present extensive simulation results showing that the error is robust with respect to the choice of states of a given rank, the random selection of settings, and that the number of settings can be significantly reduced with only a negligible increase in error. We present an argument to explain these findings based on a concentration inequality for the Fisher information matrix. In the more general setup of random basis measurements we use this argument to show that for certain rank $r$ states it suffices to measure in $O(r\log d)$ bases to achieve the average Fisher information over all bases. 
We present numerical evidence for states upto 8 atoms, supporting a conjecture on a lower bound for the Fisher information which, if true, would imply a similar behaviour in the case of Pauli bases. 
The relation to similar problems in compressed sensing is also discussed.
\end{abstract}


\maketitle

\section{Introduction}
Recent years have witnessed great experimental progress in the study and control of individual quantum systems \cite{HarocheRaimond,WisemanMilburn}. A common feature of many experiments is the use of quantum state tomography (QST) methods as a 
key tool for validating the results \cite{MIT,tenqubit}.  The aim of QST is to statistically reconstruct an unknown state from the outcomes of repeated measurements performed on identical copies of the state. Among the proposed estimation methods we mention,  e.g. variations of maximum likelihood \cite{MLparis,MLrobin,MLmofqubits,HradilML,LocalML}, linear inversion \cite{VogelRisken}, Bayesian inference \cite{BayRDgill,Baykalman}, estimation with incomplete measurements \cite{Maxentropy,Hradilincomplete,maxlikandentropy}, and continuous variables tomography  \cite{LvovskyRaymer}. 

However, for composite systems such as trapped ions, full state tomography becomes  challenging due to the exponential increase in dimension \cite{14qubit}. Therefore, there has  been a significant interest in developing tomography methods that are efficient for certain \emph{lower dimensional families of physically relevant states}. For instance, the estimation of low rank states has been considered in the context of compressed sensing (CS) \cite{CSerrorbounds,CSnoRIP,alexandra,alexandragross,CSKalev}, model selection \cite{GutaKypraiosDryden}, and spectral thresholding \cite{rankpenalised,spectralthresholding}. The estimation of the permutationally invariant part of the density matrix as an approximation to the true state is also relevant in certain physical models \cite{TothGross,TobiasToth,SchwemmerToth}. Similarly, the estimation of matrix product states  \cite{Cramer:2010} is particularly relevant for many-body systems, but also for estimating dynamical parameters of open systems \cite{CatanaGutaBouten,GutaKiukas}.

In this paper we build on the fruitful CS  idea that the sparsity of low rank states can be exploited  in order to identify and estimate the state with a \emph{reduced number of `measurements'}, in contrast to standard, informationally complete QST. 
Recall that a rank-$r$ joint state of $n$ qubits can be characterised by $O(rd)$ parameters, where $d= 2^n$ is the dimension of the associated Hilbert space. In the original CS proposal it is shown that such a state can be recovered from the expectation values of 
$O(rd \log{d})$ randomly chosen Pauli observables. More recent work concentrates on error bounds \cite{alexandra,CSerrorbounds} and confidence intervals \cite{alexandragross} of  CS estimators.  

%

However, from a statistical and experimental viewpoint the estimation based on Pauli expectations does not make the most efficient use of the measurement data available in ion trap experiments. Indeed, the Pauli expectations can be seen as `coarse grained' statistics of the `raw data' which consists of counts for individual outcomes of a measurement in an orthonormal basis. 
This coarse graining leads to  loss of information and a significant increase in estimation error, as shown in section \ref{sec.coarse}.  

In contrast, here we consider the statistical problem of estimating low rank states in the set-up of multiple ions tomography (MIT), where the input is the counts dataset provided by the experiment. The goal is to investigate the possibility of using a \emph{reduced number of measurement settings} (Pauli bases), without a significant loss of statistical accuracy, in comparison to standard, full settings MIT. For this, we consider the behaviour of the mean square error (MSE) with respect to the Frobenius distance between the true state and the estimator $\mathbb{E}\| \hat{\rho}- \rho\|_2^2$, in the limit of large number of measurement samples. 
According to asymptotic theory \cite{YoungSmith}, in this regime the MSE of efficient estimators (e.g. maximum likelihood) $\hat{\rho}$ takes the following expression
\begin{equation} \label{eq.mse.asymptotic}
\mathbb{E} \| \hat{\rho}-\rho \|_2^2 =   \frac{1}{N} {\rm Tr}(I(\rho|\mathcal{S})^{-1} G) + o(N^{-1}).  
\end{equation}
Above, $I(\rho|\mathcal{S})$ is the classical Fisher information associated with the chosen measurement design $\mathcal{S}$ and a local parametrisation of rank-$r$ states, 
and $G$ is the positive weight matrix associated with the quadratic approximation of the Frobenius distance in the local parameters. 

\begin{figure}[h]
\centering
\includegraphics[scale=0.55]{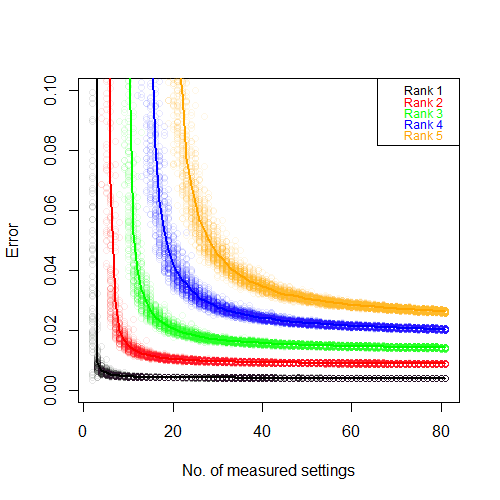}
 \caption{ Asymptotic MSE for $4$ ions states with ranks $r= 1,\ldots, 5$, and total sample size $N= m\cdot 3^4= 8100$. For each rank we chose 10 random states, and for each number of settings $k$ (horizontal axis) we chose 10 random sets of settings $\mathcal{S}$. The MSE for each such combination is represented with a circle, and the lines are the average values. }
\label{fig:settings}
\end{figure}

In the following section we review the MIT set-up, and formulate the `reduced settings hypothesis' in statistical terms. After this, we present the results of extensive numerical simulations testing this hypothesis, which are summarised in  Figure \ref{fig:settings}. We find that the asymptotic MSE given by (\ref{eq.mse.asymptotic}) is very robust with respect to a reduction the number of settings, with a choice of random settings making up the measurement design $\mathcal{S}$. For instance, 4 ions states of rank 3 can be estimated by using 20 settings (out of a total of 81) with a negligible increase in estimation error. Also, to test the validity of the asymptotic theory for low rank states, we compared the theoretical prediction (\ref{eq.mse.asymptotic}) with the actual MSE of the maximum likelihood estimator and found a very good agreement for $m=100$ repetitions per setting, a typical value used in experiments \cite{MIT}.  

To explain the observed robustness, we outline an argument based on a 
concentration inequality \cite{winter} for the Fisher information matrix of an experiment with randomly chosen Pauli settings. 
Transforming the argument into a mathematical proof requires control over certain spectral properties of the Fisher information matrix, and remains an open problem. However, by relaxing the Pauli measurement setup, and allowing for measurements with respect to \emph{random bases}, we can prove that states of rank $r$ can be estimated by using $O(r \log d)$ settings, with only a small increase in the MSE, relative to the setup in which a large number of settings is probed, cf. Theorem \ref{th.concentration}. For Pauli measurements we present numerical evidence on the lowest eigenvalue of the Fisher information matrix, supporting the conjecture that the MSE of low rank states concentrates for a small number of measurement settings.

From a CS viewpoint, our question is closely related to the work \cite{rankonemeasurements,stablelowrank} inspired by the PhaseLift problem \cite{phaseliftCandes,phaseliftKeung} which considers the case where the incomplete `measurements' are expectations of rank-one projections sampled randomly from a Gaussian distribution, or  a projective t-design.  
In \cite{onblowrank} the analysis is extended to the physically relevant case of random orthonormal basis measurements, and it is shown that rank-$r$ states become identifiable with a large probability for only $O(r \log^3{d})$ `sufficiently random' measurements. These results are in broad agreement with our findings, calling for a better understanding of the connections between the CS estimators and statistical approaches considered here.

\section{Multiple Ions Tomography with Incomplete Measurements}
In this paper we consider the multiple ions tomographic (MIT) setup as in the ion-trap experiments \cite{MIT}. In MIT the goal is to statistically reconstruct the joint state of $n$ ions (modelled as two-level systems), from counts data generated by performing a large number of measurements on identically prepared systems. The unknown state $\rho$ is a $d \times d$ density matrix (complex, positive trace-one matrix) where $d = 2^n$ is the dimension of the Hilbert space of $n$ ions. The experimenter can measure an arbitrary Pauli observable $\sigma_x,\sigma_y$ or $\sigma_z$ of each ion, simultaneously on all $n$ ions. Thus, each measurement setting is labelled by a sequence $\mathbf{s} = \left( s_1, \ldots , s_n \right) \in \{x, y, z\}^n$ out of $3^n$ possible choices. The measurement produces an outcome $\mathbf{o} = \left( o_1, \ldots ,o_n \right) \in \mathcal{O}_n := \{ +1, -1 \}^n$, whose probability is 
\begin{equation} \label{eq:prob}
p_\rho(\mathbf{o} \vert \mathbf{s} ) := \Tr(\rho P_\mathbf{o}^\mathbf{s}) = \langle \lambda_\mathbf{o}^\mathbf{s} \vert \rho \vert \lambda_\mathbf{o}^\mathbf{s} \rangle ,
\end{equation}
where $P_\mathbf{o}^\mathbf{s}$ is the one dimensional projection 
\begin{math}
P_\mathbf{o}^\mathbf{s} = \vert \lambda^{s_1}_{o_1} \rangle \langle \lambda^{s_1}_{o_1} \vert \otimes \ldots \otimes \vert \lambda^{s_n}_{o_n} \rangle \langle \lambda^{s_n}_{o_n} \vert
\end{math}, and, $ \vert \lambda^{s}_{\pm} \rangle$ is an eigenvector of the Pauli matrix $\sigma_s$, with a corresponding $\pm 1$ eigenvalue.

The measurement procedure and statistical model can be summarised as follows. For each setting $\mathbf{s}$ the experimenter performs $m$ repeated measurements and collects the counts of different outcomes $N(\mathbf{o}\vert \mathbf{s})$, so that the total number of quantum samples used is $N := m \times 3^n$. The resulting dataset is a $2^n \times 3^n$ table whose columns are independent and contain all the counts in a given setting. The overall measurement is informationally complete, and the state can be estimated by using a number of methods proposed in the literature \cite{HradilML,spectralthresholding,rankpenalised}.

Now, there are several reasons to consider a set-up in which a \emph{smaller} number of measurement settings are used for estimating the state; switching measurement settings may be more costly than repeating a measurement in the same setting, and smaller datasets may be easier to handle computationally. However, by removing even a single setting, the state becomes unidentifiable. This is because the corresponding tensor of Pauli operators is linearly independent from all the one dimensional projections of the remaining settings, and therefore its expectation value cannot be estimated. This can be remedied if some prior information about the state is available. The relevant example here is from compressed sensing \cite{CSerrorbounds,CSnoRIP,alexandra,alexandragross,CSKalev} which shows that low rank states are uniquely determined by the Pauli expectations associated with a reduced number of settings. However, the existing compressed sensing literature does not address the statistical problem of estimating the state directly from the \emph{raw} measurement data (i.e. the counts $N({\bf o}|{\bf s})$), as it typically employs \emph{coarse grained} statistics such as Pauli expectations. Our goal is to investigate the statistical efficiency of estimating low rank states with reduced measurement settings. We will consider an asymptotic scenario in which the number $m$ of measurement 
repetitions per setting is large and the mean square error can be characterised in terms of the classical 
Fisher information, as discussed above. As we show below this regime is already attained for $m=100$, which is of the order of repetitions cycles used in experiments \cite{MIT}.

As stated above, we assume that the prepared state $\rho$ belongs to the space  of rank $r$ states $\mathbb{S}_r\subset M(\mathbb{C}^d)$, for a fixed rank $r<d$. Since the asymptotic mean square error depends only on the local properties of the statistical model, it suffices to consider a parametrisation $\theta\to \rho_\theta$ of a neighbourhood of $\rho$ in 
$\mathbb{S}_r$, which can be chosen as follows. In its own eigenbasis $\rho$ is the diagonal matrix of eigenvalues
${\rm Diag}(\lambda_1, \dots, \lambda_r, 0,\dots, 0)$, and any sufficiently close state is uniquely determined by its matrix elements in the first $r$ rows (or columns). Intuitively this can be understood by noting that any rank-r state $\rho^\prime$ in the neighbourhood of 
$\rho$ can be obtained by perturbing the eigenvalues and performing a small rotation of the eigenbasis; in the first order of approximations these transformation leave the $(d-r)\times (d-r)$ lower-right corner unchanged so 
$$
\rho^{\prime} =
\left(
\begin{array}{ccc}
{\rm Diag}(\lambda_1, \dots, \lambda_r) && 0\\
&&\\
0 &&0
\end{array}
\right)
+ 
 \left(
\begin{array}{ccc}
\Delta_{diag} && \Delta_{off} \\
&&\\
\Delta^{\dagger}_{off}  && O(\|\Delta\|^2 )
\end{array}
\right).
$$
We therefore choose the (local) parametrisation $\rho^\prime = \rho_\theta$ with
\begin{eqnarray}
\theta &:= \left( \theta^{(d)}; \theta^{(r)} ;\theta^{(i)} \right) \label{eq.parametrisation} \\
&= ( \rho^\prime_{2,2}, \ldots, \rho^\prime_{r,r}\ ;  {\rm Re}\rho^\prime_{1,2}, \ldots, {\rm Re}\rho^\prime_{r,d}; {\rm Im}\rho^\prime_{1,2}, \ldots, {\rm Im}\rho^\prime_{r,d} ) \in \mathbb{R}^{2rd -r^2-1} \nonumber
\end{eqnarray}
where, in order to enforce a trace-one normalisation, we constrain the first diagonal matrix element to be $\rho^\prime_{1,1}= 1- \sum_{i=2}^{d} \rho^\prime_{i,i}$. In this parametrisation we denote $\rho= \rho_{\theta_0}$, with $\theta_0 := \left( \lambda_1,\ldots,\lambda_r; 0\ldots0 ; 0\ldots0 \right)$. The Frobenius distance is locally quadratic in $\theta$ so that 
$$
\Vert \rho_{\theta_1} - \rho_{\theta_2} \Vert^2_2 =  (\theta_1-\theta_2)^{T} G (\theta_1-\theta_2) + o(\|\theta_1-\theta_2\|^2)
$$
where 
$
G_{a,b}= {\rm Tr}\left[ \frac{\partial \rho_{\theta}}{\partial \theta_{a}} \cdot \frac{\partial \rho_{\theta}}{\partial \theta_{b}} \right]
$
is a constant weight matrix whose expression can be found in the appendix, below equation (\ref{eq.G}). After fixing the parametrisation, we now define the statistical model of multiple ions tomography with incomplete settings. Let $\mathcal{S}\subset \{x,y,z\}^n$ be a set of $k$ randomly chosen settings, and consider the modified scenario in which ions prepared in the unknown state $\rho$ are repeatedly measured $m=N/k$ times for each setting in $\mathcal{S}$, so that the overall number of quantum samples is always $N$. The classical Fisher information associated with a single chosen setting $\mathbf{s}$ is defined as 
\begin{equation}\label{eqn:Fisher}
I(\rho \vert \mathbf{s})_{a,b} :=  \sum_{\mathbf{o}} \frac{1}{p_{\rho}(\mathbf{o}\vert\mathbf{s})}\frac{\partial p_{\rho}(\mathbf{o}\vert \mathbf{s})}{\partial \theta_{a}} \cdot \frac{\partial p_{\rho}(\mathbf{o}\vert \mathbf{s})}{\partial \theta_{b}}.
\end{equation}

\noindent For a set $\mathcal{S}$ of $k$ settings the Fisher information matrix associated with a single measurement sample from each setting $\mathbf{s}\in \mathcal{S}$ is given by the sum of the individual Fisher matrices $I(\rho \vert \mathbf{s})$, and for later purposes we will denote the average
$
I(\rho \vert \mathcal{S}) = \frac{1}{k} \sum_{\mathbf{s} \in \mathcal{S}} I(\rho \vert \mathbf{s}).
$
The individual matricies can be computed by using definition (\ref{eqn:Fisher}) together with equation (\ref{eq:prob}) and the parametrisation (\ref{eq.parametrisation}).

Since the outcomes from $m$ repeated measurements in a setting $\mathbf{s}$ are i.i.d, when the number of repetitions $m$ is sufficiently large, efficient estimators  of $\theta$ (and hence of $\rho$) from these outcomes have an asymptotically Gaussian 
distribution \cite{YoungSmith}
$$
\sqrt{m} (\hat{\theta}-\theta) \approx N(0, I(\rho \vert \mathcal{S})^{-1})
$$

\noindent where the covariance matrix $I(\rho|\mathcal{S})^{-1}$ is the Fisher information associated with a single measurement sample of the set $\mathcal{S}$. From this behaviour and the local expansion of the Frobenius distance, we see that for (reasonably) large $m$, the mean square error of an efficient estimator (e.g. maximum likelihood) scales as   
\begin{equation}\label{eq.mse.fisher}
{\rm MSE}:= \mathbb{E}(\Vert \hat{\rho} - \rho \Vert^2_2) \approx \frac{1}{N} {\rm Tr}( I(\rho \vert \mathcal{S})^{-1}G).
\end{equation}


The trace expression is a measure of the sensitivity of the chosen set of settings $\mathcal{S}$ at $\rho$.  Since the settings are chosen randomly we need to study the fluctuations of ${\rm Tr}(I(\rho \vert \mathcal{S})^{-1}G)$. In the next section we present extensive simulation results which essentially show that one can significantly reduce the number of settings without affecting the MSE.

\section{Numerical Simulations}
In Figure \ref{fig:settings} we plot the values of the asymptotic MSE ${\rm Tr}(I(\rho|\mathcal{S})^{-1}G)/N$ for various ranks, choices of 4 ions states, and choices of measurement designs $\mathcal{S}$ (sets of settings).  For each rank $r=1,\dots, 5$ we generated $10$ states by using the Cholesky decomposition 
$\rho = T^{*}T$, cf. \cite{Bhatia}. For each state, the MSE values are calculated over a range of measurements with reduced settings, starting from the `full' measurement with $3^n$ settings, as follows. For a given number of reduced measurements $k$, we generated 10 independent sets $\mathcal{S}$ of randomly chosen settings. For each pair $(\rho,\mathcal{S})$  we evaluated the Fisher information $I(\rho \vert \mathcal{S})$ and the weight matrix $G$ in the parametrisation described above. In these simulations, the total number of copies of the state is kept constant as a resource. Therefore, a smaller number of measurement settings 
$k$ leads to a larger number of repetitions $m= N/k$ per setting. The simulations show that asymptotic risk for low rank states demonstrates only a gradual increase even over a significant reduction in the number of settings measured. For example, for states of rank 3, one can reduce the number of settings from 81 to 20 with a negligible increase in the MSE. Moreover, for a given $k$, 
the fluctuations of the MSE over choices of states and settings are rather small, showing the robustness of the procedure.

\begin{figure}[t]
\centering
\includegraphics[scale=0.55]{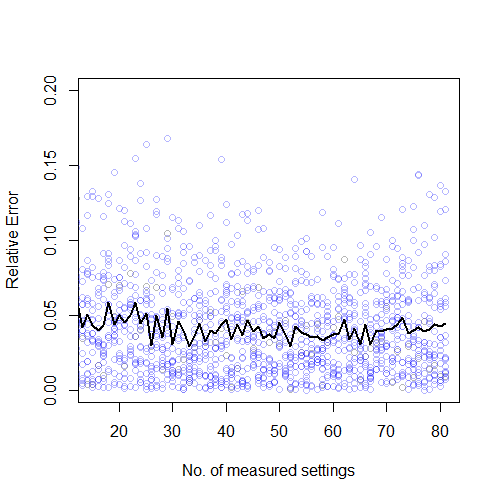}
 \caption{ The error of the (estimated) MSE of the Maximum Likelihood estimate relative to the asymptotic Fisher MSE (see \ref{eqn:MLre}), for a $4$ ion pure state. The circles plot the RE for the $20$ random sets $\mathcal{S}$ chosen for each $k$ number of  settings, and the line plots the average relative error. The total sample size is $N=8100$.}
\label{fig:ML}
\end{figure}

In the previous section we argued that the theoretical value (\ref{eq.mse.fisher}) is close to the actual error of an efficient estimator, when the number of samples is reasonably large. To verify this we have computed the  \textit{Maximum Likelihood} (ML) estimator and studied its performance in this reduced measurement settings MIT setup. The ML estimator implemented is a modified form of the iterative $R\rho R$ method in \cite{HradilML} - for the estimates generated at each iteration of the algorithm, only the $r$ largest eigenvalues are retained. This modification ensures that the ML estimator has knowledge about the rank of the density matrix. 
The results of the comparison with the Fisher prediction (\ref{eq.mse.fisher}) are shown in Figure \ref{fig:ML}, and show a very good agreement between the two. 
For a given random set $\mathcal{S}$ of $k$ settings, the MSE of the MLE $\mathbb{E}\Vert \hat{\rho}_{ML}-\rho \Vert^2_2$ is estimated by averaging the square error over $30$ ML estimates. The relative error 
\begin{equation} \label{eqn:MLre}
\left| 1-\frac{N \cdot\mathbb{E}\Vert \hat{\rho}_{ML}-\rho \Vert^2_2}{{\rm Tr}(I(\rho|\mathcal{S})^{-1}G)} \right| 
\end{equation}
is then plotted for each choice of $\mathcal{S}$ as a single circle. On average, the relative error is of order of $5\%$.  
In conclusion, the simulations indicate that low rank states can be estimated with a significantly smaller number of measurement settings than the total of $3^n$ currently used in experiments, with a negligible loss of statistical accuracy.

\section{A Concentration Bound for the MSE}

Why is the  MSE robust with respect to the reduction of the number of measured settings? 
In this section we provide an intuitive explanation based on a concentration bound  for the asymptotic MSE, i.e. 
the random function $\mathcal{S}\mapsto {\rm Tr}(I(\rho|\mathcal{S})^{-1}G)$. 
Analysing the observed MSE concentration for MIT with Pauli measurements is difficult due to the special, discrete set of bases which contribute to the average. Much like the problem of proving the the RIP property in compressed sensing \cite{yikai,CSerrorbounds}, it is easier to analyse a more random set-up, namely one where the measurement bases making up the design $\mathcal{S}$ are drawn randomly from the \emph{uniform measure} over orthonormal bases (ONB). We therefore begin by considering this general setup of random measurements and return to the Pauli measurements later in the section.

Physically, this random setup could be implemented by first rotating the state $\rho$ by a random unitary 
$U$, after which each atom is measured in the $\sigma_z$ eigenbasis. We therefore let $\mathcal{S} = \{ \mathbf{s}_1, \ldots, \mathbf{s}_k \}$ be the altered design with randomly, uniformly distributed measurement bases. Since the settings in $\mathcal{S}$ are independent, the Fisher information matrices $I(\rho|{\bf s})$ are independent, and for $k$ large enough the  average information per setting approaches the mean information over all random settings
$$
I(\rho|\mathcal{S})= \frac{1}{k} \sum_{{\bf s}\in \mathcal{S}} I(\rho | {\bf s}) \approx \bar{I}:= \mathbb{E}_{\mathbf{s}}\left[ I(\rho | {\bf s}) \right].
$$


\noindent Since we are interested in the behaviour of the MSE for the randomly chosen designs $\mathcal{S}$, we look at the relative error
\begin{equation}\label{eq.re}
RE(\rho|\mathcal{S}):= {\rm Tr}(I(\rho|\mathcal{S})^{-1}G) / {\rm Tr}(\bar{I}^{-1}G).
\end{equation}
and would like to determine the number of settings $k$ required for the MSE to be concentrated close the optimal value of $\Tr(\bar{I}^{-1}G)$.

To investigate this MSE concentration for states of rank $r$ in this setup, we focus our attention on states with equal eigenvalues, i.e. \begin{math} \rho_0 := {\rm Diag} \left( \frac{1}{r}, \ldots, \frac{1}{r},0,\ldots,0 \right) \end{math}, with respect to its eigenbasis; due to the unitary symmetry of the random settings design, the eigenbasis can be chosen to be the standard basis. The reason for choosing this particular spectrum is that such states represent the `least sparse' state of rank $r$. Indeed, rank-r states which have some eigenvalues close to zero can be approximated by states of lower ranks, and we expect that they require even smaller number of measurement settings. 
The following Theorem shows that in order to keep the relative error (\ref{eq.re}) close to 1 it suffices to take a number of random settings $k$ which scales as $O(r\log (2rd))$ with respect to rank and Hilbert space dimension. Taking into account that one setting provides $d$ probabilities, the total number of expectations is of the order $O(r d \log (2rd))$ which roughly agrees with the number of Pauli expectations required in compressed sensing. We will come back to this comparison in the following section.

\begin{theorem}\label{th.concentration}
Let $\mathcal{S}= \{{\bf s}_1, \dots, {\bf s}_k\} $ be a design with randomly, uniformly distributed measurement bases. 
Let $I_\mathcal{S}:= I(\rho_0|\mathcal{S})$ be the associated Fisher information, and let $\overline{I}$ be the mean 
Fisher information over all possible bases, both calculated at $\rho_0$ (as defined above). For a sufficiently small $\epsilon \geq 0$, the following inequality holds
$$
(1-\epsilon)  {\rm Tr}\left[\overline{I}^{-1}  G \right] \leq {\rm Tr} \left[ I_\mathcal{S}^{-1} G\right] 
\leq (1+\epsilon) {\rm Tr} \left[ \overline{I}^{-1} G \right]  \nonumber
$$
with probability $1-\delta$, provided that the number of measurements performed is $k= C(r+1) \log(2D/\delta)$, 
with $D= 2rd-r^2-1$ the dimension of the space of rank-r states.   
\end{theorem}

The proof of this theorem is detailed in the appendix, and uses a matrix Chernoff bound \cite{winter}, to bound the deviation of $G^{-1/2}I_{\mathcal{S}}G^{-1/2}$ from the mean  $G^{-1/2} \overline{I} G^{-1/2}$. This is then recast in terms of a bound on the MSE as in the theorem above. The two bounds show that with probability $1-\delta$, the relative error $RE(\rho_0|\mathcal{S})$ is in the interval $[1-\epsilon, 1+\epsilon]$, so using design $\mathcal{S}$ induces at most an $\epsilon$ increase of MSE.  Similar results can be derived along the lines of the proof, for states with arbitrary spectrum.

\begin{figure}[t]
 \centering
  \includegraphics[scale=0.55]{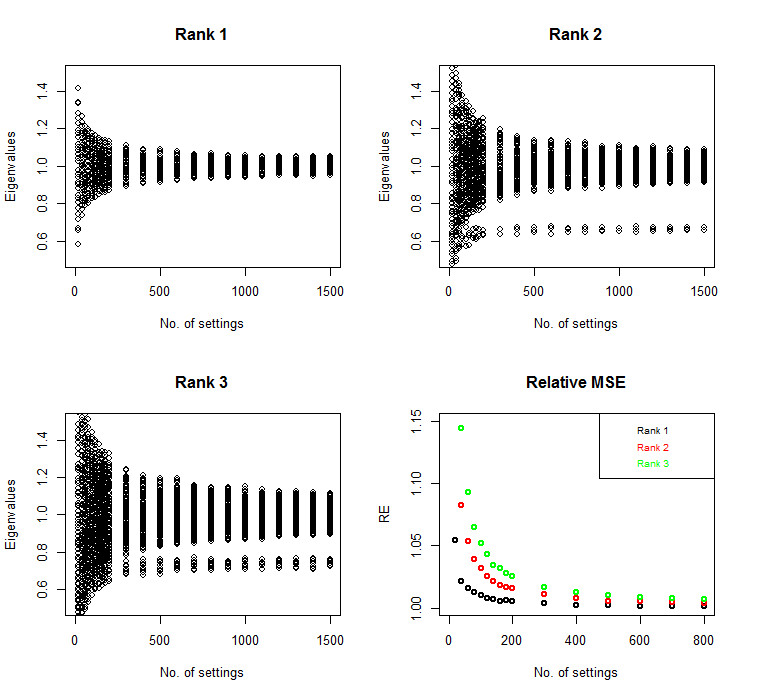}
  \caption{Plots of the eigenvalues of $G^{-1/2}I_{\mathcal{S}}G^{-1/2}$ for $k$ random settings. For a given rank $r$, we chose a random 4 ion state $\rho_0$ with $r$ equal eigenvalues. We observe a concentration of the eigenvalues as $I_{\mathcal{S}}$ approximates the mean information $\overline{I}$ (See Thm.\ref{th.concentration}). The bottom right plot graphs the relative error $RE(\rho_0 \vert \mathcal{S})$ for the different ranks.   
}
\label{fig:concentration}
\end{figure}

Figure \ref{fig:concentration} illustrates this concentration in two ways; by plotting the relative error $RE(\rho_0 \vert \mathcal{S})$ and by plotting the eigenvalues of $G^{-1/2}I_{\mathcal{S}}G^{-1/2}$, for various values of $k$. The concentration in the spectrum of the eigenvalues demonstrates the rate at which $I_{\mathcal{S}}$ approximates the mean Fisher information $\overline{I}$. We see that for pure states all eigenvalues concentrate around 1, this is because $G^{-1/2} \overline{I} G^{-1/2}$ is an identity matrix for pure states. While for ranks 2 and 3, this matrix is no longer identity and has eigenvalues that are either 1 or $r/(r+1)$. We see in the plots for these ranks that the lower band in the eigenvalues spectrum approaches the minimum eigenvalue of $r/(r+1)$, while the remaining eigenvalues concentrate around 1. The explicit form of the $G^{-1/2} \overline{I} G^{-1/2}$ matrix is detailed in the appendix. 

The above theorem guarantees that for a $4$ ions pure state, the MSE is within 5\% of the optimal, with a probability of failure $\delta = 0.1$, provided that we measure $k \approx 7100$ settings. However, the bottom-right plot in Figure \ref{fig:concentration} shows that the MSE concentrates much earlier, well within $k=100$ settings. This indicates that studying the concentration of $G^{-1/2}I_{\mathcal{S}}G^{-1/2}$ to bound the MSE provides a highly pessimistic estimate for $k$. Note however, that although the value $k \approx 7100$  is much larger than the full set of measurements for a 4 ions state in the MIT setup, the theorem demonstrates a significant reduction in the number of settings needed when we consider larger states of $n\geq 9$ ions.

\subsection{Pauli settings}
 
\begin{figure}[t]
\centering
  \includegraphics[scale=0.55]{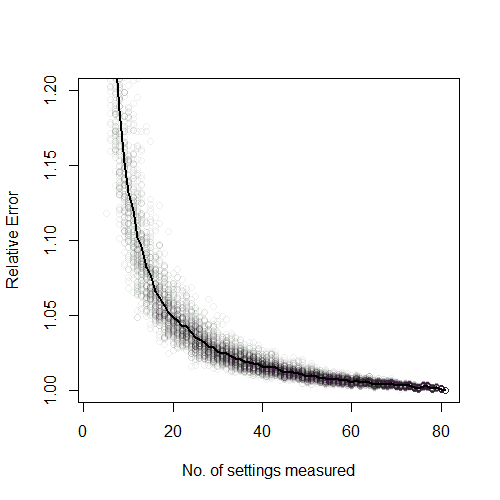}
  \caption{Relative Error $RE(\rho|\mathcal{S})$ for the Pauli settings, for a randomly chosen $4$ ion pure state. The thick line plots the mean relative error over 100 different choices of $k$ settings, with the light grey circles plotting the relative errors for different choices of the settings. 
  }
  \label{fig:RE}
\end{figure}

We now return to the more physical set-up in which the settings are chosen from the set $\{x,y,z\}^n$ of Pauli measurements. Figure \ref{fig:RE} plots the error $RE(\rho\vert \mathcal{S})$ of the MSEs for the reduced settings, relative to the MSE of the average information for full $3^n$ settings $\bar{I} = (3^n)^{-1} \sum_{\mathbf{s} \in \{x,y,z\}^n} I(\rho \vert \mathbf{s})$. The numerical simulations show that even for $k=20$ settings, the average MSE is only 5\% higher than the MSE of the full settings experiment, while when the variance is taken into account, most MSEs are less that  10\% higher. We note that in the simulations, the different settings making up the measurement design $\mathcal{S}$ are chosen without replacement, while an application of the concentration bound in the theorem would use a slightly altered setup in which the different settings are chosen independently and with equal probabilities (drawing with replacement). For a discussion on the relation between the two set-ups we refer to \cite{GrossVincent}.

The key step in establishing a concentration bound as in Theorem \ref{th.concentration} is to control the ratio 
$$
\frac{\lambda_{max}}{\lambda_{min}}:= 
\frac{\max_{\mathbf{s}} \lambda_{max}( G^{-1/2}I (\rho_0 \vert \mathbf{s}) G^{-1/2})}{\lambda_{min}( G^{-1/2}\bar{I} G^{-1/2})}
$$
between the largest maximum eigenvalue of $G^{-1/2}I (\rho_0 \vert \mathbf{s}) G^{-1/2}$ over all measurements and the minimum eigenvalue of
 $G^{-1/2}\bar{I} G^{-1/2}$. In the case of the uniformly distributed settings, $\bar{I}$ can be computed explicitly by using analytic expressions for moments of random unitaries \cite{Collins}, which gives $\lambda_{min}  =\frac{r}{r+1} $ for $r>1$, and $\lambda_{min}=1$ for pure states, while 
 $\lambda_{max}$ can be upper bounded by using the inequality between the quantum and classical Fisher informations \cite{BraunsteinCaves}, as $\lambda_{max}\leq  2 r$ for $r>1$ and $\lambda_{max}\leq 4$ for $r=1$. Together these give a $\frac{\lambda_{max}}{\lambda_{min}}= 2(r+1)$ which determines the number of measurement settings $k$ in Theorem \ref{th.concentration}.

\begin{figure}[t]
 \centering
  \includegraphics[scale=0.55]{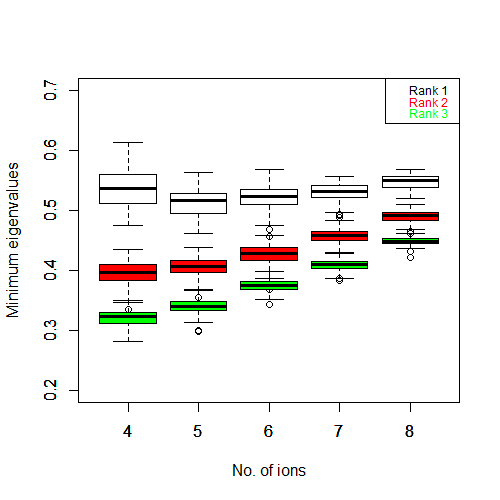}
  \caption{ Boxplots of the minimum eigenvalues of $G^{-1/2}\overline{I}G^{-1/2}$ for the Pauli settings. For a given rank and ion number, we chose randomly 100 different states $\rho_0$ with $r$ equal eigenvalues.
  }
  \label{fig:mineigenvalues}
\end{figure}

For the Pauli measurements set-up, the same upper bound holds for the maximum eigenvalue, but at the moment we do not have a similar lower bound for  $\lambda_{min}( G^{-1/2}\bar{I} G^{-1/2})$, where $\bar{I}$ is the average Fisher information over \emph{Pauli settings}. However, there is strong numerical evidence that the smallest eigenvalue of $\lambda_{min}( G^{-1/2}\bar{I} G^{-1/2})$ remains well bounded away from zero. Figure \ref{fig:mineigenvalues} plots the minimum eigenvalues for 100 states of 4 to 8 ions, over three different ranks. We notice that the minimum eigenvalue for each rank is well concentrated away from zero and for ranks $r>1$ clearly demonstrates an increase with the dimension of the space. While the full dependence of $\lambda_{min}$ on $r$ and $d$ is unclear, we conjecture that for any fixed rank, $\lambda_{min}$ is larger than a fixed constant for all states of rank $r$, of arbitrarily many ions. If this was true, it would imply that a state of {\it fixed} rank $r$ can be estimated efficiently with $O(\log d)$ settings. 


For now, as a step in the direction of proving the conjectured concentration as in Theorem \ref{th.concentration} for reduced settings, we will prove a weaker result based on 
a rough lower bound for $\lambda_{min}$. From Theorem 2 in \cite{spectralthresholding} we have that for full $3^n$ settings, the MSE of an optimal estimator $\hat{\rho}$ is upper bounded as
$$
 \mathbb{E}\Vert \hat{\rho} - \rho \Vert^2_F \leq C^{\prime} \frac{rd}{N} \log{(2d)},
$$
 
\noindent with $C^{\prime} > 0$ being an absolute constant. Asymptotically, the MSE is lower bounded by \small$1/N \cdot \Tr(\overline{I}^{-1}G)$ \normalsize  which implies
 \begin{math}
 1/\lambda_{min} \leq C^{\prime} rd \log{(2d)}.
 \end{math}
 This gives us a rough lower bound on the minimum eigenvalue. Plugging this value into the concentration bound of Theorem \ref{th.concentration} gives us that the minimum number of settings $k$ scales as $O(r^2 d \log^2{(2D}))$, which despite being far from optimal, demonstrates a better scaling than the $3^n$ of the `full settings' setup.  
 
 \section{Coarse vs Fine Grained Models}
 \label{sec.coarse}
 As mentioned in the introduction, a similar reduction in the number of `measurements' has been found in compressed sensing (CS) estimators \cite{CSerrorbounds,CSnoRIP,alexandra,alexandragross,CSKalev}, which use $O(rd \log{d})$ expectations of Pauli operators to recover the unknown state. CS techniques provide computationally efficient estimators whose estimation errors scale optimally with the number of parameters and with the errors in the estimation of the Pauli expectations. 
However, from the statistical viewpoint the Pauli expectations are not the most efficient starting point in estimation, as they are `coarse grained' statistics of the `raw', or `fine grained' measurement data given by the counts $N({\bf o}|{\bf s})$.

\begin{figure}[t]
\centering
\includegraphics[scale=0.55]{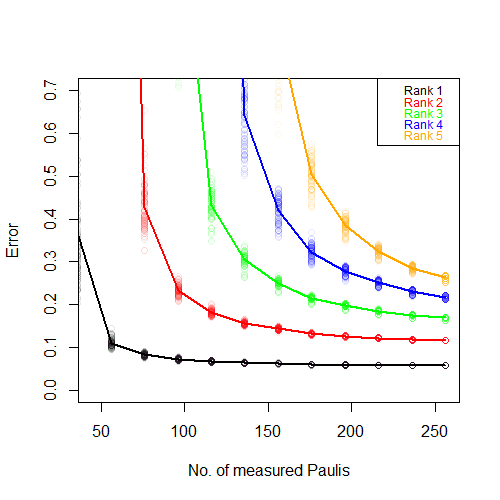}
\caption{The plot of $\Tr(I^{-1}G)$ for the `coarse grained' model for multiple ranks, with $4$ ions and $256$ total measured Paulis $\mathbf{b}$. The total number of copies of the states is 8100. The experiment is repeated over 10 different states, and 10 random choices of Pauli measurements for each state. }
\label{fig:CS}
\end{figure}

A single measurement in the `coarse grained' model is defined by a Pauli observable $\sigma_{\mathbf{b}} := \left(\sigma_{b_1}\otimes \ldots \otimes \sigma_{b_n} \right)$ with $b_i \in \{0,x,y,z \}$, where $\sigma_{0}$ is the 
identity matrix. To compute its expectation one needs to measure $\sigma_{\bf b}$ to obtain a binary outcome $\{\pm 1\}$ and average over the results. The outcomes probabilities are $p(\pm 1|{\bf b})= {\rm Tr} (\rho P^{\bf b}_{\pm})$ where $P^{\bf b}_{\pm} $ are the two spectral projections of $\sigma_{\bf b}$, and the Fisher information of this model can be computed in much the same way as that of the Pauli bases measurements.  In Figure \ref{fig:CS} the asymptotic MSE ${\rm Tr}(I(\rho \vert \mathcal{B})^{-1}G)/N$ is plotted for different sets of randomly chosen Pauli observables $\mathcal{B}:= \{{\bf b}_1, \dots, {\bf b}_k\}$. On comparison with Figure \ref{fig:settings} we see that minimum number of measurements that need to measured in the Pauli bases model is much smaller. Additionally, the risk for a full set of measurements is an order of magnitude larger in the `coarse grained' model. This increase in the asymptotic risk has also been pointed out in \cite{GutaKypraiosDryden}. 

The discrepancy can be explained by noting that the measurement of $\sigma_{\bf b}$ is a coarse graining of a finer, ONB measurement of a setting such that $s_i= b_i$ whenever $b_i\neq 0$. Indeed, using the spectral decomposition of $ \sigma_{\bf b}$,  
we can compute its expectation as
$$
\langle \sigma_{\bf b}\rangle =  {\rm Tr} (\rho\sigma_{\bf b}) 
=\sum_{{\bf o}} 
\left(\prod_{i : b_i\neq 0} o_i\right) ~p({\bf o} | {\bf s}). 
$$
By replacing the probabilities $p({\bf o} | {\bf s})$ in the above formula by the empirical frequencies $N({\bf o}|{\bf s})/m$ we obtain the estimate of the Pauli expectations. However, by constructing this statistic we loose a large amount of information contained in the frequencies, which explains the increase in the MSE.

%

 \section{Conclusions}
In this paper we investigated the statistical performance of reduced settings measurements in ion tomography. We did not focus on a particular estimation method but rather on how the accuracy of efficient estimators (which achieve the asymptotic scaling (\ref{eq.mse.asymptotic}) of the MSE) depends on the state and the measurement design. We found that for low rank states, the experimenter can measure a small proportion of randomly selected settings without a significant increase in the MSE. Furthermore we presented a possible line of argument for a mathematical proof based 
on concentration inequality for the Fisher information. In the case of measurements with respect to random bases we showed that certain states of rank $r$ can be estimated with $O(r\log d)$ settings with an $\epsilon$ increase in MSE compared with designs with a large number of settings. It remains an open question whether the same scaling of the size of the measurement design holds for the Pauli measurements, but we presented strong numerical evidence that the Fisher information may satisfy the required spectral properties.

In a future work we plan to apply these ideas and construct estimators in a more realistic setup where the rank is not a priori known.

\ack
We thank Andreas Winter for bringing to our attention the matrix concentration inequality, during the  mini-workshop ``Mathematical Physics meets Sparse Recovery", at Mathematisches Forschungsinstitut Oberwolfach.  We also thank 
David Gross for organising the mini-workshop and for many stimulating discussions. This work was supported by the EPSRC grant EP/J009776/1.

\section*{References}


\newpage

\appendix
\section*{Appendix}

\section{Proof of Theorem 1}
\begin{proof}[\nopunct]
\noindent As briefly mentioned in the main text of the paper, the proof of the theorem utilises the following matrix Chernoff bound \cite{winter}, where the random matrices $X_i$ are given by $G^{-1/2}I(\rho_0 \vert \mathbf{s}_i) G^{-1/2}$, with 
${\bf s}_i$ random bases.

\begin{theorem}{(\textbf{Matrix Chernoff})} Consider a finite sequence $X_1,\ldots,X_k$ of independent, random, positive matrices with dimension $D$, such that $\lambda_{max}(X) \leq R$. For $\mathbb{E}X = M \geq \mu \mathbf{1}$ and $0 \leq \epsilon \leq \frac{1}{2}$, 
\begin{equation}
\mathbb{P} \left\{ \frac{1}{k} \sum_{i=1}^{k} X_i \not\in \biggl[ (1-\epsilon)M, (1+\epsilon)M \biggl] \right\} 
 \leq 2D \cdot \exp{\biggl(-k \cdot \frac{\epsilon^2 \mu}{2R \cdot \log{2}} \biggl)}
\end{equation}
\end{theorem}

\noindent We note that $G^{-1/2}I_{\mathcal{S}}G^{-1/2}$ is a sum of $k$ independent, random, positive matrices. In order to apply the above bound, we need to upper bound the largest eigenvalue of $G^{-1/2}I(\rho_0 \vert \mathbf{s}) G^{-1/2}$ over all measurements. We also need to lower bound the smallest eigenvalue of the expected Fisher information $G^{-1/2}\overline{I}(\rho_0)G^{-1/2}$. We will first derive these bounds and then derive the result by applying the Chernoff bound.

As in the text, we work with the local parametrisation 
\begin{equation}
\theta = \left( \theta^{(d)}, \theta^{(r)}, \theta^{(i)} \right) = \left( \rho_{2,2}, \ldots, \rho_{r,r}; Re \rho_{1,2}, \ldots, Re \rho_{r,d}; Im \rho_{1,2}, \ldots, Im \rho_{r,d} \right) \nonumber
\end{equation}

\noindent where $\rho_{1,1}$ is constrained to enforce the trace-one normalisation. The Fisher information therefore, has the following block structure
\begin{equation*}
I(\rho)=\left(
\begin{array}{ccccc}
I^{dd}(\rho) && I^{dr}(\rho) && I^{di}(\rho) \\ 
&&&&\\
I^{rd}(\rho) && I^{rr}(\rho) && I^{ri}(\rho) \\ 
&&&&\\
I^{id}(\rho) && I^{ir}(\rho) && I^{ii} (\rho)
\end{array}\right)
\end{equation*}

\noindent with the superscripts identifying the parameters considered; diagonal, real and imaginary. The weight matrix $G$ also has the same block structure with elements  
\begin{equation}\label{eq.G}
G_{a,b}= \Tr\left[ \frac{\partial \rho_{\theta}}{\partial \theta_{a}} \cdot \frac{\partial \rho_{\theta}}{\partial \theta_{b}} \right]
\end{equation}

In the parametrisation described above, the weight matrix $G$ has the following block diagonal form:
\begin{enumerate}
\item The \textit{diagonal-diagonal} block:
\begin{enumerate}
\item $G^{dd}_{a,b} = 1+ \delta_{a,b}$
\end{enumerate}
\item The \textit{real-real} and \textit{imaginary-imaginary} block:
\begin{enumerate}
\item $G^{rr/ii}_{a,b} = 2 \cdot \delta_{a,b}$
\end{enumerate}
\end{enumerate}

\noindent with the other blocks being zero. We note that both the Fisher, and the weight matrix are of dimension $D:=2rd-r^2-1$. 
\\

\noindent \textit{\textbf{Bound on the largest eigenvalue}}---We use the inequality $I(\rho_0 \vert \mathbf{s} ) \leq F$ between the classical and quantum Fisher informations to bound the largest eigenvalue of $G^{-1/2}I(\rho_0 \vert \mathbf{s} ) G^{-1/2}$ over all measurements by the largest eigenvalue of $G^{-1/2}F(\rho_0)G^{-1/2}$. 
The derivation of the quantum Fisher matrix presented here follows \cite{QFI}. We calculate the quantum Fisher information in the local parametrisation described above, and evaluate it at the diagonal state $\rho_0$. 

We begin by considering a state $\rho_{\theta}$ locally around some arbitrary rank-$r$ state $\rho^{\prime}$, and write the spectral decomposition as:
\begin{equation}
\rho_{\theta} = \sum_{i=1}^{r} p_i \vert \psi_i \rangle \langle \psi_i \vert
\end{equation}
The quantum Fisher information matrix is defined as:
\begin{equation} \label{eqn:QFI}
F_{a,b} = \Tr\left[ \rho_{\theta} (L^a_\theta \circ L^b_\theta) \right] = \frac{1}{2} 
\Tr \biggl[ \rho_{\theta} \left( L^a_\theta L^b_\theta + L^b_\theta L^a_\theta \right) \biggl]
\end{equation}
where the symmetric logarithmic derivatives are defined through the equation:
\begin{equation}
\partial_{\theta_a} \rho_{\theta} = L^a_\theta \circ \rho_{\theta} = \frac{1}{2} \left( L^a_\theta \rho_{\theta} + \rho_{\theta} L^a_\theta \right)
\end{equation}

\noindent We determine the elements of this matrix in the ONB formed by the eigenbasis set $\{ \vert \psi_i \rangle \}$ 
\begin{eqnarray} \label{eqn:SLD}
\langle \psi_i \vert \partial_{\theta_a} \rho_{\theta} \vert \psi_j \rangle &=  \frac{1}{2} \langle \psi_i \vert L^a_\theta \rho_{\theta} \vert \psi_j \rangle +  \frac{1}{2} \langle \psi_i \vert \rho_{\theta} L^a_\theta \vert \psi_j \rangle \nonumber\\
&= \frac{1}{2} (p_j + p_i) \langle \psi_i \vert L^a_\theta \vert \psi_j \rangle 
\end{eqnarray}
As pointed out in \cite{QFI}, $L^a_\theta$ (and $L^b_\theta$) is in principle supported on the full space, but its entries for $i,j > r$ are arbitrary. However, the Fisher information does not use values for which $i,j >r$. This can be seen by expanding (\ref{eqn:QFI}) in the following way 
\begin{eqnarray}
F_{a,b} &= \frac{1}{2} \Tr \left[ \sum_i p_i \vert \psi_i \rangle \langle \psi_i \vert (L^a_\theta L^b_\theta) + \sum_i p_i \vert \psi_i \rangle \langle \psi_i \vert (L^b_\theta L^a_\theta) \right] \nonumber\\ 
&= \frac{1}{2} \sum_i^r \sum_j^d p_i \left( L^a_{\theta;i,j} L^b_{\theta;j,i} + L^b_{\theta;i,j} L^a_{\theta;j,i} \right) \nonumber
\end{eqnarray}
Since the index $i\leq r$, (\ref{eqn:SLD}) can be inverted inside the expansion of the Fisher information as 
\begin{equation}
L^a_{\theta;i,j} = \frac{2(\partial_{\theta_a} \rho_{\theta})_{i,j}}{p_i+p_j}
\end{equation}
The quantum Fisher matrix therefore becomes 
$$
F_{a,b} = \sum_i^r \sum_j^d \frac{4p_i}{(p_i + p_j)^2}  Re \biggl[ (\partial_{\theta_a} \rho_{\theta})_{i,j} (\partial_{\theta_b} \rho_{\theta})_{j,i} \biggl] 
$$
where we used the fact that $\partial_{\theta_{a/b}} \rho_{\theta}$ is self-adjoint. Since $\rho_{\theta}$ is parameterised by its matrix elements in the eigenbasis $\{ \vert \lambda_i \rangle \}$ of the state $\rho^{\prime}$, we can use this to write the partial derivatives out explicitly. Using the notation that $r_a, c_a$ represents the row and column indices for the parameter $\theta_{a}$, the quantum Fisher matrix now becomes:

\begin{eqnarray}
 F_{a,b}^{d,d} =   \sum_i^r \sum_j^d \frac{4p_i}{(p_i + p_j)^2} Re \Biggl[ &\langle \psi_i \vert \biggl( \vert \lambda_{r_a} \rangle \langle \lambda_{r_a} \vert - \vert \lambda_1\rangle \langle \lambda_{1} \vert \biggl) \vert \psi_j \rangle \nonumber\\
 & \langle \psi_j \vert \biggl( \vert \lambda_{r_b} \rangle \langle \lambda_{r_b} \vert - \vert \lambda_1\rangle \langle \lambda_1 \vert \biggl) \vert \psi_i \rangle \Biggl] \nonumber
\end{eqnarray}
 for the \textbf{diagonal-diagonal} block, and for the rest
\begin{eqnarray}
F_{a,b} = \sum_i^r \sum_j^d \frac{4p_i}{(p_i + p_j)^2}  Re \Biggl[ &\langle \psi_i \vert \biggl( \vert \lambda_{r_a} \rangle \langle \lambda_{c_a} \vert + \vert \lambda_{c_a} \rangle \langle \lambda_{r_a} \vert \biggl) \vert \psi_j \rangle \nonumber \\
& \langle \psi_j \vert \biggl( \vert \lambda_{r_b} \rangle \langle \lambda_{c_b} \vert +\vert \lambda_{c_b} \rangle \langle \lambda_{r_b} \vert \biggl) \vert \psi_i \rangle \Biggl] \nonumber
\end{eqnarray}
We now evaluate these last two equations at $\theta = \theta_{0}$, for our special state that is diagonal with entries given by $1/r$. At this state $\vert \psi_i \rangle = \vert \lambda_i \rangle$.  The \textit{diagonal-diagonal} block of the Fisher matrix has elements:
 \begin{equation}
\left. F^{d,d}_{a,b} \right|_{\theta = \theta_{0}} = r \left( 1+ \delta_{r_a,r_b} \right) 
 \end{equation}
\noindent While the \textit{real-real} and \textit{imaginary-imaginary} blocks are diagonal with elements
\begin{equation}
\left. F_{a,b} \right|_{\theta = \theta_{0}} = \frac{4}{p_{r_a}+p_{c_a}} \left( \delta_{r_a, r_b} \cdot \delta_{c_a, c_b} \right)
\end{equation}
It is easy to see that the \textit{real-diagonal, identity-diagonal} blocks are all zero. The \textit{real-imaginary} blocks are zero since we consider only $Re\left[ (\partial_{\theta_a} \rho_{\theta})_{i,m} (\partial_{\theta_b} \rho_{\theta})_{m,i} \right]$. Therefore, the elements of the quantum Fisher matrix are:
\begin{enumerate}
\item For the \textbf{\textit{Diagonal-Diagonal}} block with $r>1$,
\begin{enumerate}

\item  $\left. F^{dd}_{a,a} \right|_{\theta = \theta_{0}}= 2r ~$when $r_a \leq r$
\item $\left. F^{dd}_{a,b} \right|_{\theta = \theta_{0}}= r$  when $r_a, r_b \leq r$, and $a \neq b$
\end{enumerate}

\item For the \textbf{\textit{Real-Real}} and \textbf{\textit{Imaginary-Imaginary}} blocks:

\begin{enumerate}
\item $\left. F^{rr/ii}_{a,a}\right|_{\theta = \theta_{0}}= 2r$ when $r_a < c_a \leq r$ 
\item  $\left. F^{rr/ii}_{a,a}\right|_{\theta = \theta_{0}}= 4r$ when $r_a \leq r, c_a > r$
\end{enumerate}

\end{enumerate}

\noindent On comparing this with the weight matrix $G$, we notice that both $G$ and $F$ have the same block diagonal structure, with the off-diagonal blocks being zero. So we can write

$$
G^{-1/2}FG^{-1/2} = {G^{dd}}^{-1/2}F^{dd}{G^{dd}}^{-1/2} \bigoplus {G^{rr}}^{-1/2}F^{rr}{G^{rr}}^{-1/2} 
\bigoplus {G^{ii}}^{-1/2}F^{ii}{G^{ii}}^{-1/2} 
$$

We notice that $F^{dd} = r \cdot G^{dd}$,  which gives us
$$
G^{-1/2}FG^{-1/2} = r \mathbf{1}_{(r-1)} \bigoplus \frac{1}{2} F^{rr} \bigoplus \frac{1}{2} F^{ii} 
$$

\noindent The maximum eigenvalue of this matrix comes from the diagonal block matrices $F^{rr/ii}/2$, and is $2r$ for $r>1$, and $4$ for $r=1$. 
\\

\noindent \textit{\textbf{Bound on the smallest eigenvalue}}--- We are now interested in evaluating the smallest eigenvalue of the average Fisher information $G^{-1/2}\overline{I}(\rho_0) G^{-1/2}$.  As in \cite{spectralthresholding} we let 
$$
\mathbf{B}_{U} := \biggl\{ \vert \mathbf{o} ; U \rangle := U \vert \mathbf{o} \rangle : \mathbf{o} = 1, \dots, 2^n \biggl\}
$$
denote the ONB basis obtained by rotating the standard basis by a random unitary U. With this notation, we get that for randomly chosen basis 
\begin{equation}
G^{-1/2}\overline{I}(\rho_0) G^{-1/2} := G^{-1/2} \cdot \int \mu(dU) I (\rho_0 \vert \mathbf{B}_U) \cdot G^{-1/2}
\end{equation}

\noindent where $\mu(dU)$ is the Haar measure over unitaries used for generating the random basis. The integral in the above equation has been evaluated in \cite{spectralthresholding}, and we do not reproduce the calculation here. However, we point out that the integral in \cite{spectralthresholding} has been evaluated for a slightly different parametrisation of the state. Since we constrain the $\rho_{1,1}$ element, the partial derivatives in our parametrisation become
\begin{eqnarray*}
\centering
&\frac{\partial p_{\rho} (\mathbf{o} \vert \mathbf{B}_U)}{\partial \rho_{i,i}} = \vert \langle \mathbf{o} ,U \vert i \rangle \vert^2 - \vert \langle \mathbf{o},U \vert 1 \rangle \vert^2 \\
&\frac{\partial p_{\rho} (\mathbf{o} \vert \mathbf{B}_U)}{\partial Re\rho_{i,j}} = 2Re(\langle i \vert \mathbf{o}, U \rangle \langle \mathbf{o},U \vert j \rangle ) \\
& \frac{\partial p_{\rho} (\mathbf{o} \vert \mathbf{B}_U)}{\partial Im\rho_{i,j}} = 2Im(\langle i \vert \mathbf{o}, U \rangle \langle \mathbf{o},U \vert j \rangle ) 
\end{eqnarray*}

\noindent Going through the calculation with this change gives us 
\begin{enumerate}
 \item The \textbf{\textit{Diagonal-Diagonal}} block with $r>1$:
\begin{enumerate}
\item $\left. \overline{I}^{dd}_{a,a} \right|_{\theta=\theta_0} = \frac{2r}{r+1}$ when $r_a \leq r$
\item $\left. \overline{I}^{dd}_{a,b} \right|_{\theta=\theta_0} = \frac{r}{r+1}$ when $r_a, r_b \leq r$, and $a \neq b$
\end{enumerate}
\item The \textbf{\textit{Real-Real}} and \textbf{\textit{Imaginary-Imaginary}} blocks are diagonal with:
\begin{enumerate}
\item $\left. \overline{I}^{rr/ii}_{a,a} \right|_{\theta=\theta_0} = \frac{2r}{r+1}$ when $r_a < c_a \leq r$
\item $\left. \overline{I}^{rr/ii}_{a,a} \right|_{\theta=\theta_0} = 2$ when $r_a \leq r , c_a > r$ 
\end{enumerate}  
\end{enumerate}

\noindent On comparing this with the weight matrix $G$, we once again notice that both $G$ and $\overline{I}(\rho_0)$ have the same block diagonal structure, with the off-diagonal blocks being zero. So we can write
$$
G^{-1/2}\overline{I}G^{-1/2} = {G^{dd}}^{-1/2}\overline{I}^{dd}{G^{dd}}^{-1/2} \bigoplus {G^{rr}}^{-1/2}\overline{I}^{rr}{G^{rr}}^{-1/2}
\bigoplus {G^{ii}}^{-1/2}\overline{I}^{ii}{G^{ii}}^{-1/2} \nonumber
$$

We notice that $\overline{I}^{dd} = \frac{r}{r+1} \cdot G^{dd}$,  which gives us
$$
G^{-1/2}\overline{I}G^{-1/2} = \frac{r}{r+1} \mathbf{1}_{(r-1)} \bigoplus \frac{1}{2} \overline{I}^{rr} \bigoplus \frac{1}{2} \overline{I}^{ii} 
$$

\noindent The minimum eigenvalue of this matrix is $r/r+1$ for $r>1$ and $1$ for pure states.
\\

\noindent \textit{\textbf{Putting it all together}}-- We can now substitute these values into the matrix Chernoff bound. While the value of the minimum eigenvalue differs for $r>1$ and $r=1$, the final bound remains the same because the upper bounds are different in these cases. Therefore here we calculate the bound for the case when $r>1$. Writing $P_{\mathcal{S}}=G^{-1/2}I_{\mathcal{S}}G^{-1/2}$ and $\overline{P} = G^{-1/2}\overline{I}G^{-1/2}$ for notational simplicity, we have for $r>1$
$$
\mathbb{P}\left\{ P_{\mathcal{S}} \not\in \left[ (1- \epsilon) \overline{P}, (1+\epsilon) \overline{P} \right] \right\} 
\leq  2 D \cdot \exp{ \left( -k \frac{\epsilon^2}{4 (r+1) \cdot \log{2}} \right)} := \delta
$$
 
\noindent Therefore, with probability $1-\delta$ we have that 
$$
(1- \epsilon) \overline{P} \leq P_{\mathcal{S}} \leq (1+\epsilon) \overline{P}
$$

\noindent This can be re-written in the form of inequalities of Mean Square Errors with $\epsilon>0$ sufficiently small
\begin{equation*}
({1- \epsilon}) \Tr \left( \overline{P}^{-1} \right) \leq \Tr \left[  P_{\mathcal{S}}^{-1} \right] \leq (1+ \epsilon) \Tr \left( \overline{P}^{-1} \right)
\end{equation*}

For a fixed value of $\epsilon$ and $\delta$, we see that the minimum number of settings $k$ required for the above abound to hold with probability greater than $1-\delta$ is 
\begin{equation}
k = C \cdot (r+1)  \log{\left(\frac{{2D}}{\delta}\right)}  
\end{equation}

\noindent where $C := 4 (\log{2}/\epsilon^2)$ and $D := 2rd-r^2-1$.  

\end{proof}
\setcounter{section}{1}

\end{document}